\documentclass[12pt]{article}
\usepackage{latexsym}
\usepackage{float}
\usepackage{amstex}
\usepackage{portland}
\usepackage{epsfig}

\begin{document}

\title{Simulation of UHE muons propagation for GEANT3}

\author{S. Bottai 
\footnote{Address for corrispondence: 
 Dipartimento di Fisica di Firenze and INFN Sezione di Firenze, 
Largo E.Fermi n.~2, I-50125 Firenze (Italy)
tel. +39-055-229420, fax +39-055-229330} 
\\
{\it Dipartimento di Fisica, Universit\`a di Firenze and}\\ 
      {\it  INFN Sezione di Firenze}\\
\\
 L. Perrone\\
{\it Dipartimento di Fisica, Universit\`a di Lecce and}\\
     {\it INFN Sezione di Lecce}}
\date{}
\maketitle


\begin{abstract}
{A simulation package 
 for the transport of high energy muons    
has been developed. 
It has been conceived to replace  
  the muon propagation software modules implemented in 
 the detector simulation program GEANT3.
Here we discuss the results achieved  
   with our  
 package and we check the agreement with   
 numerical calculations up to $10^{8}$ $GeV$.
 }
\end{abstract}

\section{Introduction}
\label{sect1}
Underground and underwater detectors for neutrino astronomy require
simulation tools capable to correctly handle the propagation
of high energy
muons up to $PeV$ energy region and above.
The new object-oriented (C++) version of the detector simulation tool
GEANT (GEANT4 \cite{g4}), will probably be suitable for this
goal.\\
However many experiments~\cite{macro,nestor,Antares,Wiebush} 
still make use of GEANT3~\cite{manual} to simulate the detector
response.
Even if such package can in principle perform the simulation of particle
propagation above 10 $TeV$, it has been mainly designed for accelerator
experiments, whose  tipical energy range doesn't exceed
the $TeV$ region.\\
The simulation of muonic interactions is actually guaranteed by the 
 authors for muon energy below 10 $TeV$ \cite{manual}. 
 This is due to the parametrizations 
 of cross sections for radiative processes contained in GEANT3,  
   which are reliable only for energy up to 10 $TeV$. Moreover
 the description of photonuclear interaction is realized in a frame    
   which significantly disagrees with theoretical 
 calculations, for each muon energy.\\
  Our new simulation code (GMU) has been carried out to make
the GEANT3 standard library reliable in reproducing
UHE muons propagation through matter.\\
 GMU replaces the simulation of the radiative muonic interactions
performed by GEANT3, for
 each energy and for each material, keeping the same 
 structure of the original program.
 The procedure to apply the new code is
  completely transparent in such a way that no 
 adjustments 
must be implemented in the programs that make use of the standard GEANT3
library.\\

\section{Radiative Cross sections in GEANT3 and GMU}

High-energy muons propagating through matter mainly interact by 
  quasi-continuous (ionization) and discrete (bremsstrahlung,
 direct electron-positron pair production, photonuclear interaction)
 energy losses mechanisms.
Ionization dominates at energy lower than few hundreds of $GeV$
while above the $TeV$ region energy losses are mainly due to radiative
processes.\\
The radiative processes are simulated stocastically 
 by GEANT3 above a fixed (user supplied)  
 transferred energy threshold; below this threshold
 they are treated as continuous. 
The models which describe the interaction processes are
 the physical input for the code.
In more detail, simulating a given process requires:
\begin{itemize}
\item
To evaluate the probability of occurrence of the process by sampling
 the total cross section of the process
\item
To generate the final state after interaction by sampling the
  differential cross section  of the process
\end{itemize}
 The reliability of the simulation is then affected
 both by the formulas (and/or parametrizations) chosen for cross sections
     and by
 the alghoritms used for sampling and for numerical integration.\\

Standard reference formulas for UHE  calculations 
can be found in the paper by Lohmann and Voss~\cite{Lohmann} 
which also tabulate the average muon energy losses for many 
 materials and compounds up to $10$ $TeV$.\\
Direct electron pair production differential cross section
 has been first calculated by Kelner and Kotov in the framework
 of QED theory \cite{kelner}. We have used
 the well-known parametrization
 performed by Kokoulin and Petrukhin \cite{koko} which considers
 the corrections for atomic and nuclear form factors.\\ 

For bremsstrahlung differential cross section we have used the formula derived
 by Andreev and Bugaev \cite{Andreev} which
   takes into account the structure of nuclear target
 (elastic and inelastic form factors) and the exact contributions due to
   atomic electrons (screening effect and bremsstrahlung of the muons
 on electrons).
The formula used by Lohmann and Voss had
been carried out just from this one
by Petrukhin and Shestakov \cite{Shest} by neglecting
bremsstrahlung on electrons and nuclear
effects in light materials ($Z<10$). 
The discrepancy between the two approaches
reaches few percents, in terms of
bremsstrahlung energy losses, 
 in the worse case (for materials with $Z \sim 10$) and
it is almost always negligible in the calculation of the total
muon energy losses.\\

For the photonuclear interaction we have used the differential
 cross section calculated by Bezrukov and Bugaev \cite{bezrukov}
 within the vector meson dominance hypothesis. We also have considered
 the recent accelerator data coming from ZEUS and H1 \cite{Zeus}
 \cite{H1}
 according to the ref. \cite{Naumov}.
For this process we have not considered, at present, the angle between the
ingoing and outgoing muon, taking the reasonable approximation of
completely forward scattering. As a consequence, the code is not suitable for
studies dedicated to muon-nucleus scattering at large angles.\\
Total cross section
and average energy loss 
for each  radiative process $k$ can be calculated, 
 starting from differential cross sections,  as follows:\\
 \begin{equation}
 \sigma^{k}_{Tot}=\int_{vmin}^{vmax}
 \frac{d\sigma^{k}}{dv} (v,E) \,dv
\end{equation}
\begin{equation}
 -\Big\langle \frac{dE}{dx}\Big\rangle_{k}=\frac{N_{A}}{A} \, E \,
         \int_{vmin}^{vmax}\, v
 \frac{d\sigma^{k}}{dv}(v,E) \,dv
\label{losses}
\end{equation}
\begin{itemize}
\item
$N_{A}$ and $A$ are Avogadro's number and the mass number respectively 
\item
$v$ is the fraction of initial energy $E$ lost by muon 
at the occurrence of the process $k$
\item
$vmin$ and $vmax$ are the kinematical limits for the allowed values of
$v$ (\cite{bezrukov},\cite{Lohmann})
\item
$x$ is the thickness of throughgone matter, expressed in $g/cm^{2}$
\item
\[\frac{d\sigma^{k}}{dv} (v,E)\] is the differential cross section
 for the process $k$ (explicit formulas are given in the Appendix)
\end{itemize}

The calculation of energy losses  
  performed with the formulas described above 
 agree with the Lohmann and Voss results at percent level in the 
energy range they considered ($1 \, GeV$-$10 \, TeV$.)\\  
In fig.~\ref{cross} we report the total cross sections for the radiative 
  processes
as they are tabulated at the initialization procedures by standard GEANT321
and by GEANT321 plus GMU.\\

In the first case the points are calculated  with 
 approximated analytical parame\-tri\-zations whose accuracy is guaranteed
 (at least for bremsstrahlung and pair production)
 within $5 \% $ in the declared energy range ($E < 10$ $TeV$).\\
In the second case 
the points are calculated
 by performing numerical
 integration of the formulas given in the Appendix. 
(This operation requires a small amount of CPU time at the initialization,
depending also on the number of materials considered.)\\

The photonuclear cross sections differ
by more than one order of magnitude. This discrepancy
has already been observed in previous
works \cite{battist} and it is expected to affect the total energy loss
especially in the case of light materials.
The total cross sections
for bremsstrahlung and pair production below 100 $TeV$ seem to be in
excellent agreement but for energy above 100 $TeV$ the GEANT321 cross sections
 strongly deviate from expectation.\\

\section{Simulation of muon energy loss}

In order to evaluate the correctness of the simulation
code we have performed some checks for different materials.
For each test run we have activated the full stochastical regime
and we have evaluated the average energy loss resulting from the simulation.
The values achieved with standard GEANT321 and with GEANT321 plus GMU have
been compared with numerical calculations 
 of the same quantities [see eq.(\ref{losses})].\\

In fig.~\ref{water} and  
\ref{std}
we present the results for water and for standard rock respectively. 
The simulation performed by GEANT321 plus GMU reproduces very well
the numerical calculation at any energy, both for
total energy losses and for the individual processes.\\
For such materials,  
the GEANT321
simulation is in good agreement
   up to 100 $TeV$, but it produces much greater
 average energy loss for energy exceeding 100 $TeV$. \\
In table 1 we report the total and the
photonuclear 
 energy losses in hydrogen as they result from simulation performed 
 by the two codes    
($\bigtriangleup \, \Longrightarrow$  GEANT321;
$\star \, \Longrightarrow$ GEANT321+GMU). The discrepancy is
not negligible, even at low energy, being 
  around $20 \% $ at 10 $TeV$. 
The main source of this disagreement is clearly due to a  
different estimation of the photonuclear process, whose
relative importance grows for light materials.\\ 

In order to further verify the reliability of the simulation code
we performed some specific runs by recording
  the fraction $v$ of
energy lost by muon  at the occurrence of each process.
(Indeed
the average energy loss is not especially sensitive
 to very big or very small energy transfer).

 The distributions of such variable 
  have been compared with
 analytical expressions for differential cross sections.
The results are shown in fig.~\ref{sampling1}, 
\ref{sampling2} and \ref{sampling3} for a fixed muon 
 energy ($E=100$ $TeV$): they
 confirm the correctness of the sampling performed by GMU.

\section{Conclusions}

The simulation of muon propagation carried out by
standard GEANT3 seems to
be out of control for energy above $10^{5}$ $GeV$.\\
Moreover GEANT3 slightly underestimates the energy losses for
very light materials (at lower energies too), 
due to a strong  discrepancy between
expected and simulated values of the muon photonuclear cross section.\\
GEANT321 in addition to the new code GMU performs complete agreement
with expectation in the tested energy range (up to $10^{8}$ $GeV$).\\
Further improvements
will consist in a more accurate simulation
of the direction of outgoing particle produced in photonuclear
interaction and in taking into account   
the influence of the medium
(LPM effect) wich becomes important in the process
of muon bremsstrahlung at extreme high energy. In this region 
 direct muon pair production by muons
should be taken into account too.\\\\

The GMU code can be requested by conctacting the authors. E-mail:
\begin{itemize}
\item
 Bottai\verb+@+fi.infn.it
\item
 Lorenzo\verb+@+le.infn.it
\end{itemize}

\section{Acknowledgments}

We would like to thank      
  the members of the Lecce MACRO group  
   and prof. V.A.Naumov 
  for useful discussions  
 and concrete help.


\clearpage

\begin{center}
\appendix{\Large\bf{APPENDIX}}
\end{center}
\vspace{1cm}
\begin{center}
\Large\bf{Radiative cross sections formulae}
\end{center} 

\begin{itemize}
\item
$N_{A}$  
is Avogadro's number
\item
$Z$ and  
 $A$ are the atomic number and the atomic weight of the material 
\item
$v$ is the fraction of initial energy $E$ lost in the interaction 
 ($E'=E(1-v)$) 
\item
$\alpha$ is the fine structure constant
\item 
$r_{e}$
is the classical electron radius 
\item
$\lambda_{e}=r_{e} \alpha^{-1}$
is the electron Compton wavelength
\item 
$m_{e}$, $m_{\mu}$, $m_{\pi}$ and $M_{p}$  
  are the electron, muon, pion and proton rest masses ($c=1$) 
  respectively
\end{itemize}

The formulae for radiative differential cross sections are   
 taken from ref.~\cite{koko}
for $e^{+}e^{-}$ pair-production,  
from ref.~\cite{Andreev} for bremsstrahlung and 
from ref.~\cite{bezrukov} and~\cite{Naumov} 
for photonuclear interaction.

\vspace{1.5cm}

\begin{center}
$\bullet$ {\bf PAIR PRODUCTION}\\
\end{center}

\begin{equation*}
 \frac{d\sigma_{p}}{dv d\rho}= \alpha^{4} \frac{2}{3\pi} \,\,  
          (Z \lambda_{e})^{2}
          \frac{1-v}{v}
         [ \Phi_{e} + \frac{m_{e}^{2}}{m_{\mu}^{2}} \Phi_{\mu}] 
\end{equation*}

\begin{equation*}
\begin{split}
\Phi_{e} = & \{[(2+\rho^{2})(1+\beta)+\xi(3+\rho^{2})]
ln(1+\frac{1}{\xi})\\
& +\frac{1-\rho^{2}-\beta}{1+\xi}-(3+\rho^{2})\}S_{e} 
\end{split}
\end{equation*}

\begin{equation*}
\begin{split}
\Phi_{\mu}=& \{[(1+\rho^{2})(1+\frac{3\beta}{2})-\frac{1}{\xi}(1+2\beta )
(1-\rho^{2})]ln(1+\xi )\\
& +\frac{\xi (1-\rho^{2}-\beta )} {1+\xi}+
(1+2\beta)(1-\rho^{2})\}S_{\mu}
\end{split}
\end{equation*}

\begin{equation*}
\begin{split}
S_{e}  & = ln\frac{RZ^{-1/3}\sqrt{(1+\xi)(1+Y_{e})}}
{1+\frac{2m_{e}\sqrt{e}RZ^{-1/3}(1+\xi)(1+Y_{e})}{Ev(1-\rho^{2})}}\\
& -\frac{1}{2}ln[1+(\frac{3}{2} \frac{m_{e}}{m_{\mu}}Z^{1/3})^{2}
 (1+\xi)(1+Y_{e})]
\end{split}
\end{equation*}

\begin{equation*}
S_{\mu}=ln \frac{ (2/3)(m_{\mu}/m_{e})R Z^{-2/3}}
{1+\frac{2m_{e}\sqrt{e}RZ^{-1/3}(1+\xi)(1+Y_{\mu})}{Ev(1-\rho^{2})}}
\end{equation*}

\begin{equation*}
Y_{e}~=~\frac{5-\rho^{2}+4\beta(1+\rho^{2})}
{2(1+3\beta)ln(3+\frac{1}{\xi})-\rho^{2}-2\beta(2-\rho^{2})}
\end{equation*}

\begin{equation*}
Y_{\mu}~=~\frac{4+\rho^{2}+3\beta(1+\rho^{2})}
{(1+\rho^{2})(\frac{3}{2}+2\beta)ln(3+\xi)+1-\frac{3}{2}\rho^{2}} 
\end{equation*}

\begin{equation*}
\xi~=~\frac{m_{\mu}^{2} v^{2}}{4m_{e}^{2}} \frac{(1-\rho^{2})}{(1-v)};~~
~\beta~=~\frac{v^{2}}{2(1-v)};
\end{equation*}

\begin{equation*}
\rho= \frac{E^{+}-E{-}}{E^{+}+E{-}} 
\end{equation*}

The integration limits are:

\begin{equation*}
\frac{4 m_{e}}{E} \leq v \leq 1- \frac{3 \sqrt{e}}{4 E }~m_{\mu}~Z^{1/3}
\end{equation*}

\begin{equation*}
0 \leq |\rho| \leq~\left( 1-\frac{6 m_{\mu}^{2}}{E^{2}(1-v)} \right)
 \sqrt{1-\frac{4m_{e}}{Ev}}
\end{equation*}

where:\\
\begin{itemize}
\item
$E^{+}$ and $E^{-}$ are the energies of the positron and electron
\item
$R$=189 is the radiation logarithm value 
\item
$e=exp(1)$
\end{itemize}
We have   
 taken into account the influence of the atomic electrons 
  replacing $Z^{2}$ by $Z(Z+1)$.\\





\vspace{1.5cm}

\begin{center}
$\bullet$ {\bf BREMSSTRAHLUNG}\\
\end{center}

\begin{equation*}
 \frac{d \sigma_{b}}{dv} = \alpha (\frac{1}{v}) \left(2 r_{e} Z 
\frac{m_{e}}{m_{\mu}}\right)^{2} \left\{ \left(1+\frac{E^{'2}}{E^{2}}\right) 
\Phi_{1}(q_{\rm min},Z)- \frac{2}{3} 
\frac{E^{'}}{E} \Phi_{2}(q_{\rm min},Z)\right\}
\end{equation*}

\begin{equation*}
\Phi_{1,2}(q_{\rm min},Z)=\Phi^{0}_{1,2}(q_{\rm min},Z)
\end{equation*}

\begin{equation*}
\begin{split}
\Phi^0_1 (q_{\rm min},Z)=&\frac{1}{2} \left(1+\ln \frac{m^{2}_{\mu} 
a^{2}_{1}}{1+x_{1}^{2}} \right) - x_{1} \arctan \frac{1}{x_{1}}\\
    &+ \frac{1}{Z} 
\left[\frac{1}{2}\left(1+\ln \frac{m^{2}_{\mu}a_{2}^{2}}{1+x^{2}_{2}}\right)
-x_{2} \arctan\frac{1}{x_{2}}\right]
\end{split}
\end{equation*}

\begin{equation*}
\begin{split}
\Phi_{2}^{0}(q_{\rm min},Z)=& \frac{1}{2} \left(\frac{2}{3}+\ln 
 \frac{m_{\mu}^{2} 
a_{1}^{2}}{1+x_{1}^{2}} \right) +2 x_{1}^{2} \left(1-x_{1}\arctan \frac{1}{
x_{1}}+\frac{3}{4}\ln \frac{x_{1}^{2}}{1+x_{1}^{2}} \right)\\
 &+\frac{1}{Z} 
\left[ \frac{1}{2} \left( \frac{2}{3}+\ln\frac{m_{\mu}^{2}a_{2}^{2}}
{1+x_{2}^{2}} \right) +2 x_{2}^{2} \left( 1 - x_{2} 
 \arctan\frac{1}{x_{2}}+\frac{3}{4} \ln \frac{x_{2}^{2}}{1+x_{2}^{2}}
\right)\right] \;,
\end{split}
\end{equation*}

\begin{equation*}
\Delta_{1}(q_{\rm min},Z)=\ln\frac{m_{\mu}}{q_{c}}+\frac{a}{2} \ln\frac{a+1}{a-1}
\end{equation*}

\[
\Delta_{2}(q_{\rm min},Z)=\ln\displaystyle{\frac{m_\mu}{q_e}}+
\displaystyle{\frac{a}{4}}(3-a^{2})
\ln\displaystyle{\frac{a+1}{a-1}}+\displaystyle{\frac{2m_\mu ^2}{q_c^2}}
\]

\begin{equation*}
q_{\rm min} \simeq \frac{m_{\mu}^{2}v}{2E(1-v)}
, \;\;\; x_{i}=a_{i}q_{\rm min}, 
\end{equation*}

\begin{equation*}
a=\sqrt{1+\frac{4m_\mu ^2}{q_c^2}},\;\;\;\; q_{c}=1.9m_{\mu}Z^{-1/3},
\end{equation*}

\begin{equation*}
a_{1}=\frac{184.15}{\sqrt{e}Z^{1/3}m_{e}},
\;\;\; a_{2}=\frac{1194.0}{\sqrt{e}Z^{2/3}m_{e}},\;\;\;
\end{equation*}

$\bullet$  where $e=exp(1)$

The integration limits are:

\begin{equation*}
 0 \leq v \leq 1- \frac{3 \sqrt{e}}{4 E }~ m_{\mu}~Z^{1/3}
\end{equation*}

\vspace{1.5cm}

\begin{center}
$\bullet$ {\bf PHOTONUCLEAR INTERACTION}
\end{center}

\newcommand{\beq}{\begin{equation}}
\newcommand{\eeq}{\end{equation}}
\newcommand{\ea}{\end{array}}
\newcommand{\be}{\beta}
\newcommand{\de}{\delta}
\newcommand{\al}{\alpha}
\newcommand{\ga}{\gamma}
\newcommand{\ze}{\zeta}
\newcommand{\si}{\sigma}
\newcommand{\lam}{\lambda}
\newcommand{\om}{\omega}
\newcommand{\dps}{\displaystyle}
\newcommand{\la}{\label}
\newcommand{\bea}{\begin{array}}

\begin{equation*}
\begin{split}
\dps{\frac{d\si_n}{dv}} & = \; \dps{\frac{\al}{8\pi}} A\si_{\ga p}
\left(\nu\right)v \Big\{H\left(v\right)\ln \left(1+\dps{\frac{m_2^2}{t}}\right)
-\dps{\frac{2m_{\mu}^2}{t}} \\
\\
 & + \; G\left(z\right)\left[H\left(v\right)\ln
\left(1+\dps{\frac{m_1^2}{t}}\right)-H\left(v\right)\dps{\frac{2m_1^2}{m_1^2+t}}
-\dps{\frac{2m_{\mu}^2}{t}}\right] \\
\\
& + \; \dps{\frac{2\xi m_{\mu}^2}{t}}\left[G\left(z\right)\dps{\frac{2m_1^2}{m_1^2+t}} +
\dps{\frac{m_2^2}{t}}\ln \left(1+\dps{\frac{t}{m_2^2}}\right)\right] \Big\} \\
\label{photonuclear}
\end{split}
\end{equation*}
$\nu=vE$ is the energy of virtual photon 
\[
\begin{array}{c}
H(v) = 1-\dps{\frac{2}{v}}=\dps{\frac{2}{v^2}} , \;\;\;\;
G(z)=\frac{9}{z}\left\{\dps{\frac{1}{2}}+\dps{\frac{1}{z^2}}[(1+z)e^{-z}-1]\right\},\\
\\
z=0.00282 A^{\frac{1}{3}}\si_{\ga p}(\nu), \;\;\;\; t=\dps{\frac{m_{\mu}^2 v^2}
{1-v}},   \\
\\
m_1^2=0.54 \rm{GeV}^2, \;\; m_2^2=1.80 \rm{GeV}^2, \;\; \xi=0.25 \;.
\end{array}
\]

The differential cross section is proportional to the total cross section
$\sigma_{\gamma N}$, 
for absorption of a real photon of energy 
$\nu=s/2m_{N}=vE$ by a nucleon. In this calculation 
we have used the Regge-type parametrization for $\sigma_{\gamma N}$, 
 according ref.~\cite{Donna} 
This model performs the best fit to accelerator data~(\cite{Zeus},\cite{H1}). 
\begin{equation*}
\si_{\ga N} = \left[67.7 s^{0.0808} + 129 s^{-0.4525}\right] \;\mu b
\end{equation*} 
The integration limits are:
\begin{equation*}
  \frac{1}{E}(m_{\pi}+\frac{m_{\pi}^{2}}{2M_{p}}) 
 \leq 
 v \leq
1-(1+\frac{m_{\mu}^{2}}{M_{p}^{2}})\frac{M_{p}}{2E}
 \end{equation*}

\clearpage

\begin{figure}[ht]
\begin{center}
\epsfig{figure=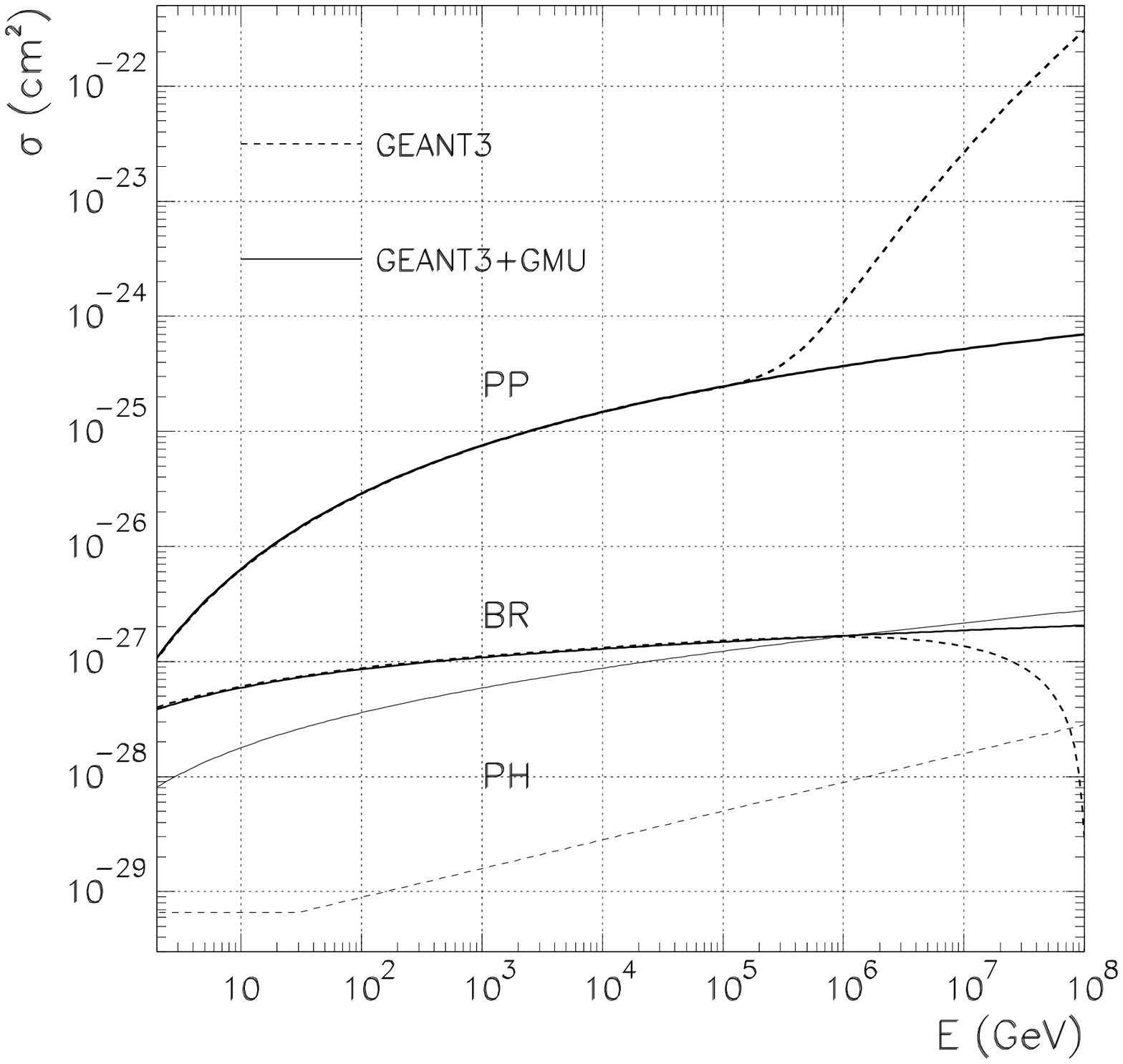,height=16cm,width=14.5cm}
\caption{}
\label{cross}
\end{center}
\end{figure}

\clearpage

\begin{figure}[ht]
\begin{center}
\epsfig{figure=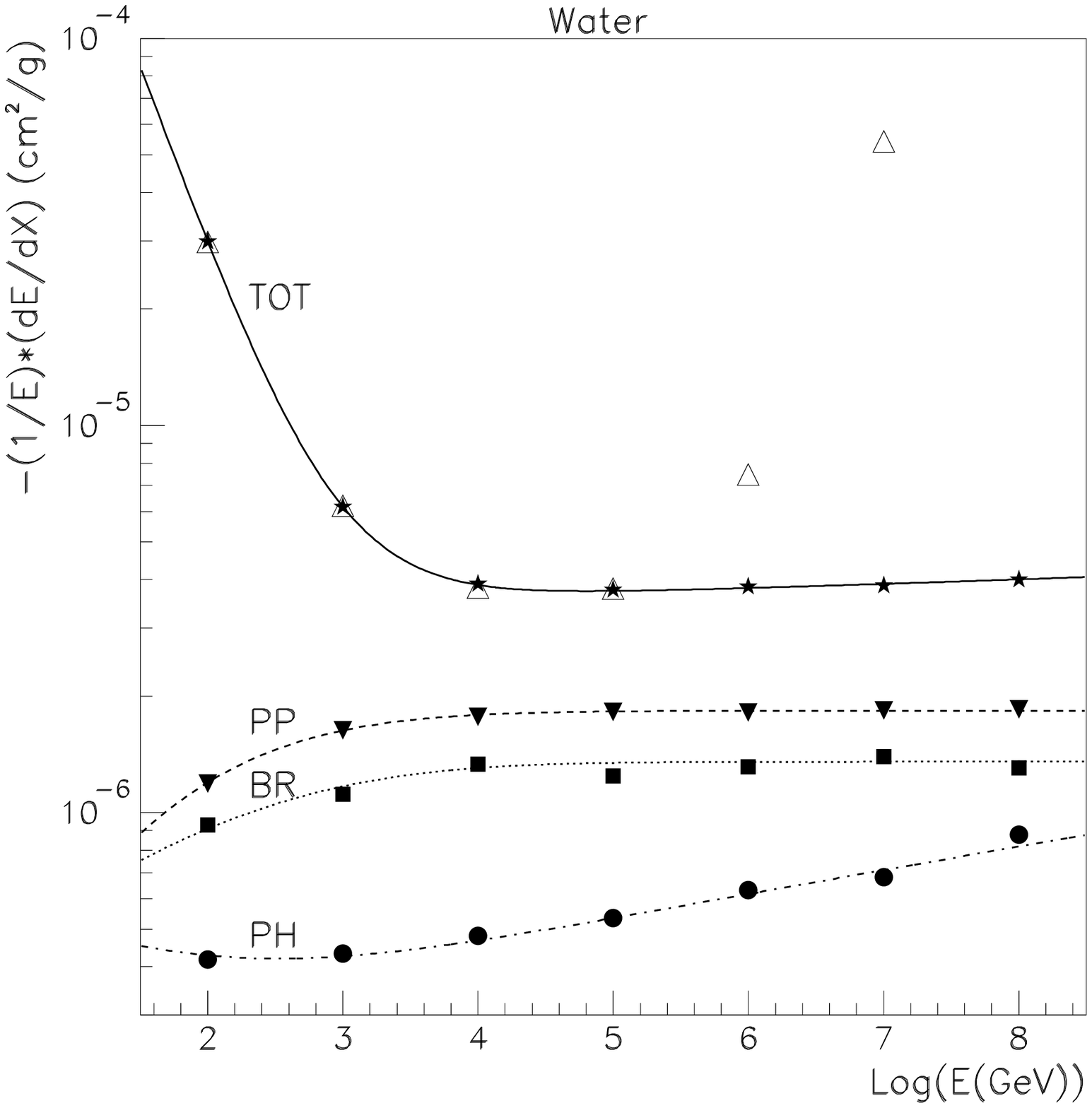,height=14.5cm,width=14.5cm}
\caption{}
\label{water}
\end{center}
\end{figure}

\clearpage

\begin{figure}[ht]
\begin{center}
\epsfig{figure=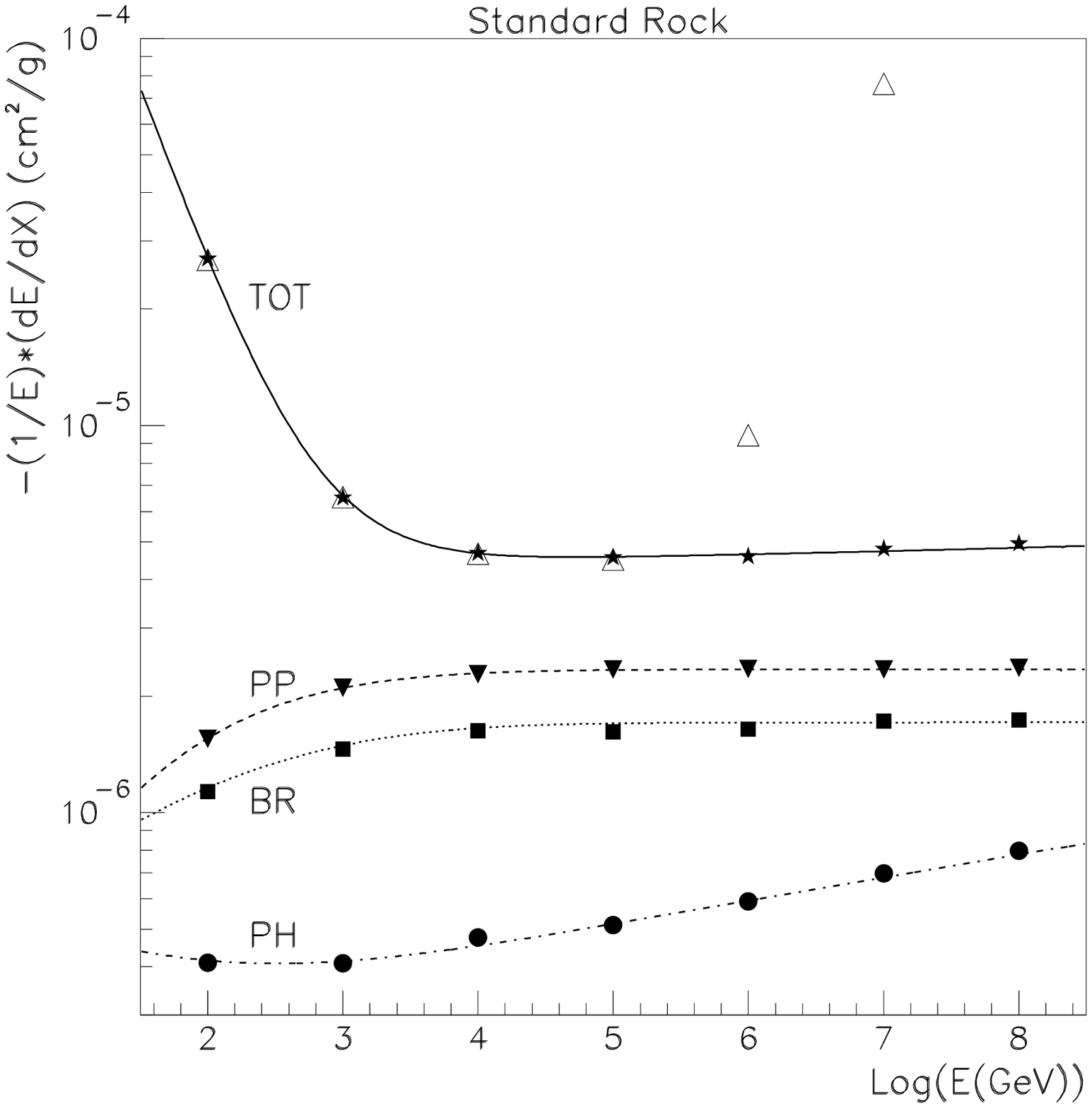,height=14.5cm,width=14.5cm}
\caption{}
\label{std}
\end{center}
\end{figure}

\clearpage

\begin{figure}[ht]
\begin{center}
\epsfig{figure=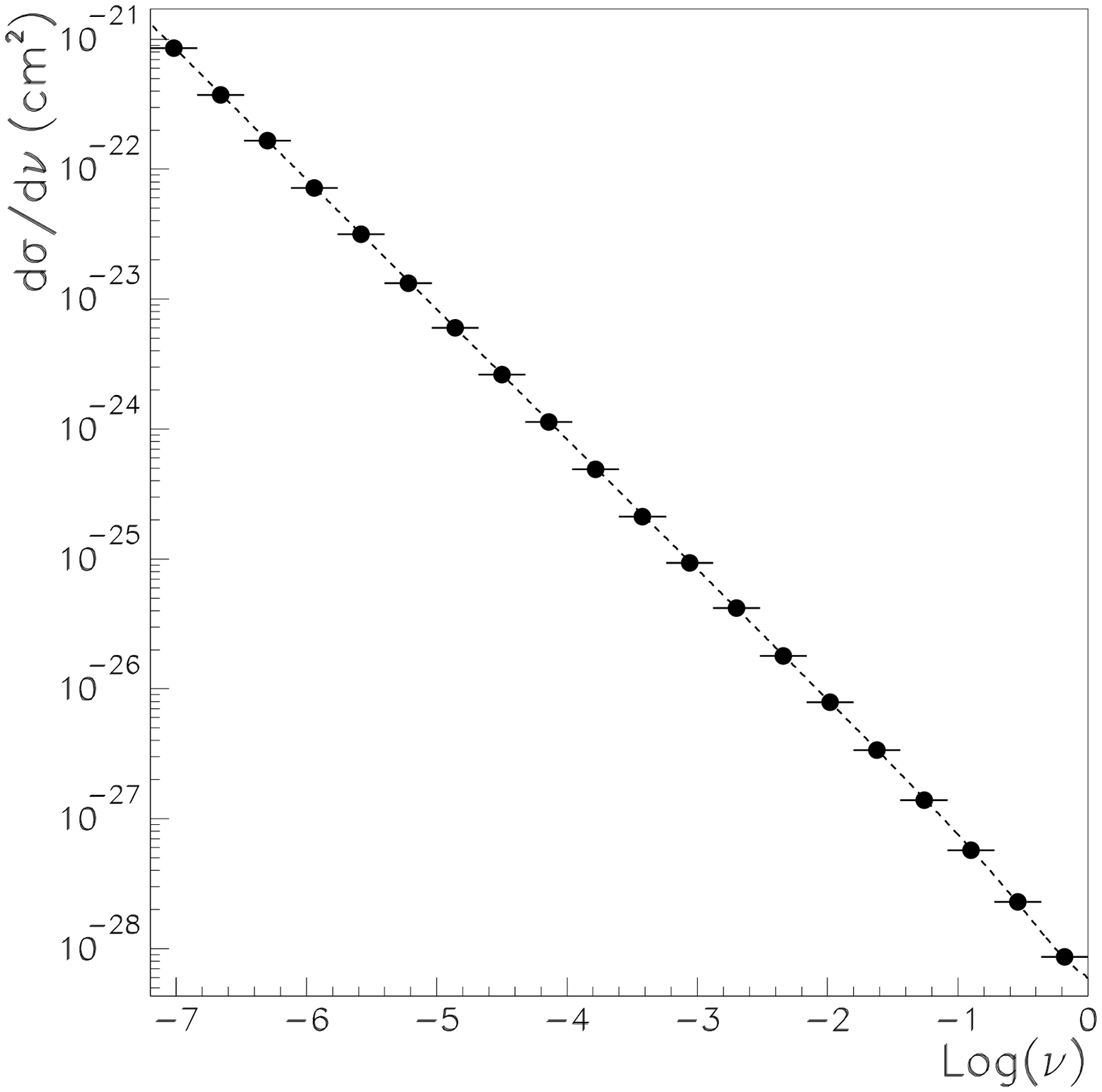,height=15cm,width=14.5cm}
\caption{}
\label{sampling1}
\end{center}
\end{figure} 

\clearpage
 
\begin{figure}[ht]
\begin{center}
\epsfig{figure=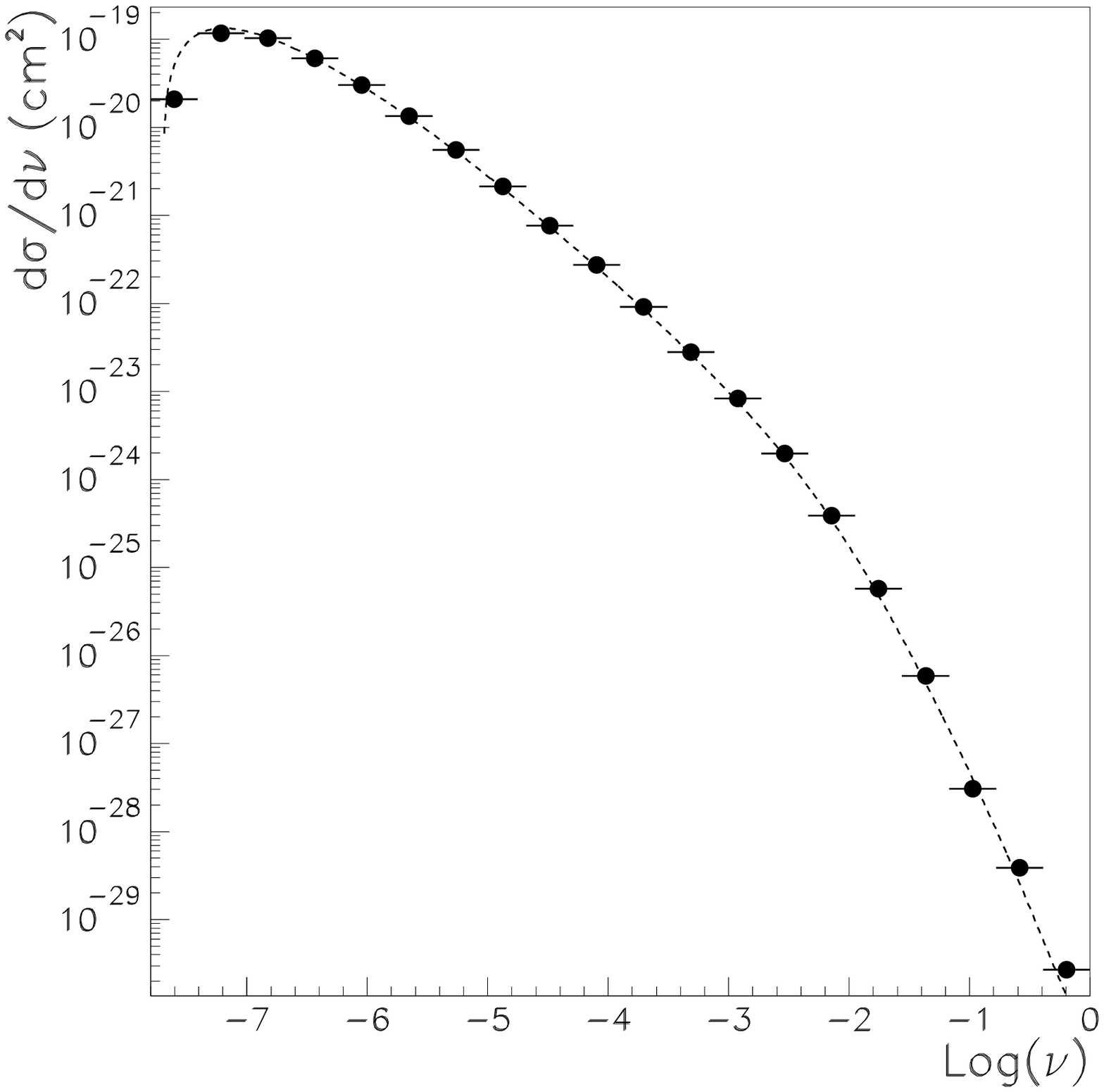,height=15cm,width=14.5cm}
\caption{}
\label{sampling2}
\end{center}
\end{figure} 

\clearpage

\begin{figure}[ht]
\begin{center}
\epsfig{figure=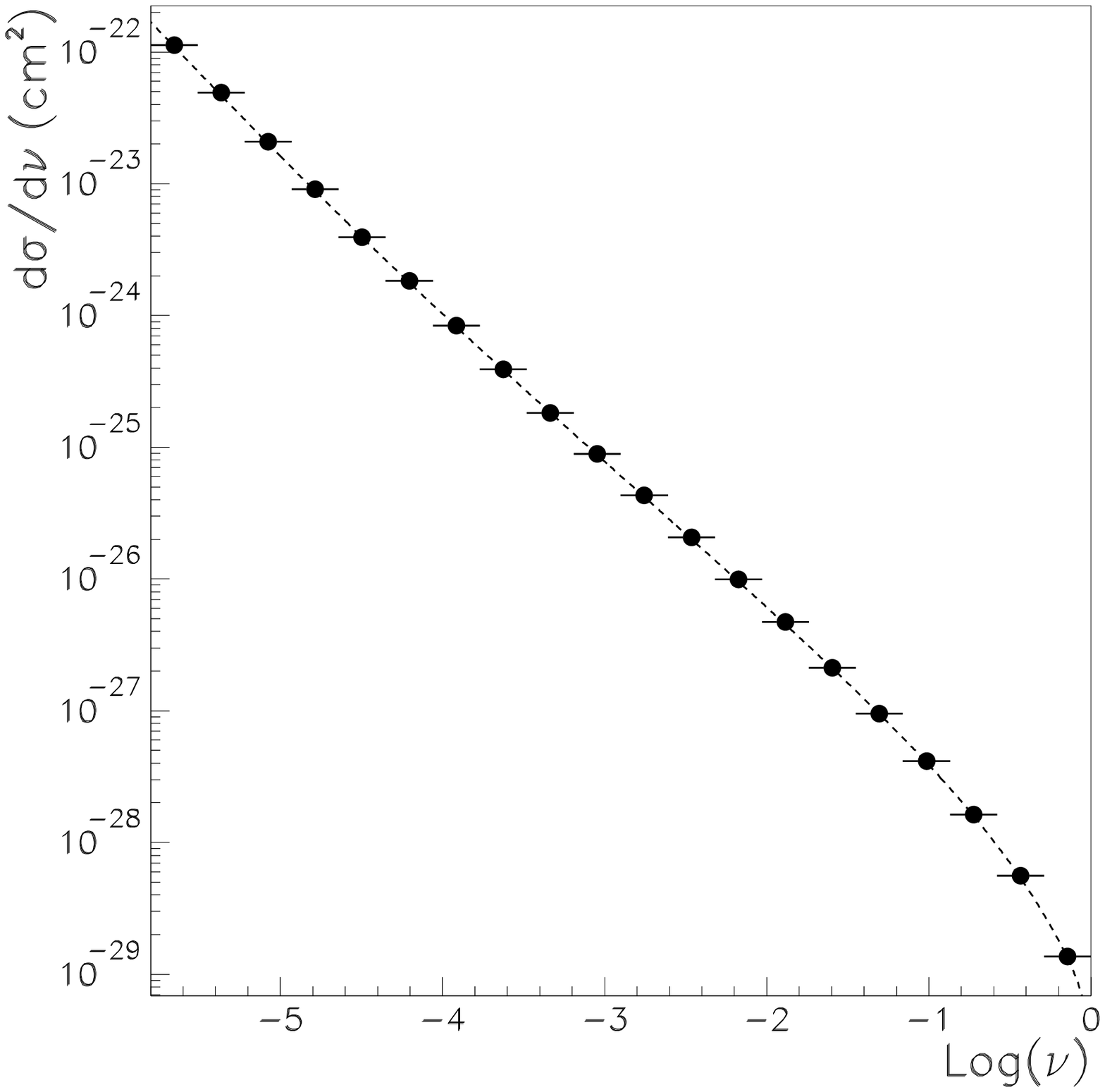,height=15cm,width=14.5cm}
\caption{}
\label{sampling3}
\end{center}
\end{figure}
\clearpage

Figure 1: 
Total Cross Sections for radiative interaction 
 processes vs muon energy (PP~$\rightarrow$ $e^{+}e^{-}$ pair-production, 
 BS~$\rightarrow$~bremsstrahlung, PH~$\rightarrow$~photonuclear interaction). 
 The plot shows the values tabulated by   
   GEANT321 (dashed lines) and by GEANT321 plus GMU (full lines).
 The material is  
   Standard Rock ($Z=11$, $A=22 $).\\

Figure 2: 
Average total muon energy loss in Water ($Z/A=0.555$) 
 vs the $Log_{10}E$ ($E$ is the muon energy). 
The plot shows   
 the results from GEANT321 (empty triangles) and from 
 GEANT321 plus GMU (full stars);   
 the contributions from individual radiative processes (performed 
 by GEANT321 plus GMU) and 
results from numerical calculations (lines) are also shown.
Statistical error bars are comparable with dots size in case of
 photonuclear interactions (PH) and bremsstrahlung (BR).\\

Figure 3: 
Average total muon energy loss in Standard Rock ($Z=11$, $A=22 $) 
 vs the $Log_{10}E$ ($E$ is the muon energy). 
The plot shows the results  
  from GEANT321 (empty triangles) and from 
 GEANT321 plus GMU (full stars);   
 the contributions from individual radiative processes (performed 
 by GEANT321 plus GMU) and 
results from numerical calculations (lines)  are also shown.
Statistical error bars are comparable with dots size in case of
 photonuclear interactions (PH) and bremsstrahlung (BR).\\

Figure 4:
Differential cross section for 
 muon bremsstrahlung vs $Log_{10}v$ ($v$ is the fraction of  
 energy $E$ lost by muon    
 in the interaction). 
The numerical calculation (dashed line)
is compared with the result obtained from the simulation
  performed with the code GEANT321 plus GMU. 
The material is Standard Rock ($E=100$ $TeV$).\\ 

Figure 5: 
Differential cross section for pair production  
vs $Log_{10}v$ ($v$ is the fraction of
 energy $E$ lost by muon
 in the interaction). 
The numerical calculation (dashed line)
is compared with the result obtained from the simulation
  performed with the code GEANT321 plus GMU. 
The material is Standard Rock ($E=100$ $TeV$).\\


Figure 6: 

Differential cross section for photonuclear interaction   
vs $Log_{10}v$ ($v$ is the fraction of
 energy $E$ lost by muon
 in the interaction). 
The numerical calculation (dashed line)
is compared with the result obtained from the simulation
  performed with the code GEANT321 plus GMU. 
The material is Standard Rock ($E=100$ $TeV$).\\

\clearpage

\begin{table}[htb]
\centering
\begin{tabular}{||c|c|c|c||}\hline\hline
E & &  $\left( 1/E \right) \cdot \left( dE/dx \right) _{Tot}$ &
  $\left( 1/E \right) \cdot \left( dE/dx \right) _{Phot}$  \\
($TeV$) & & $\left( cm^{2} \cdot g^{-1} \right)$&
      $ \left( cm^{2} \cdot g^{-1} \right)$ \\ \hline
0.1& $\bigtriangleup$ & $ \left(5.270 \pm 0.005 \right) \cdot 10^{-5}$ 
   & $ \left(0.0030 \pm 0.0001 \right) \cdot 10^{-5}$ \\ \cline{2-4}
   & $\star$ & $ \left(5.322 \pm 0.005 \right) \cdot 10^{-5}$ 
   & $ \left(0.047 \pm 0.001 \right) \cdot 10^{-5} $ \\ \hline \hline

1 & $\bigtriangleup$ & $ \left(0.704 \pm 0.004 \right) \cdot 10^{-5}$ 
  & $ \left(0.0062 \pm 0.0002 \right) \cdot 10^{-5}$ \\ \cline{2-4}
  & $\star$ & $ \left(0.738 \pm 0.004 \right)\cdot 10^{-5}$ 
  & $ \left(0.046 \pm 0.001 \right) \cdot 10^{-5} $\\ \hline \hline

10& $\bigtriangleup$ & $ \left(0.224 \pm 0.002 \right) \cdot 10^{-5} $
  & $ \left(0.0155 \pm 0.0005 \right) \cdot 10^{-5}$ \\ \cline{2-4}
  & $\star$ & $ \left(0.262 \pm 0.002 \right) \cdot 10^{-5} $ 
  & $ \left(0.0515 \pm 0.001 \right) \cdot 10^{-5} $ \\ \hline \hline
\end{tabular}
\label{tabella:tabella}
\caption{
Total energy losses for Hydrogen in comparison with
the photonuclear interaction contributions.   
 $\bigtriangleup \, \Longrightarrow$  GEANT321;
$\star \, \Longrightarrow$ GEANT321 + GMU}
\end{table}

\end{document}